\documentclass[12pt,preprint]{aastex} 
\usepackage{lineno}
\usepackage{ulem}
 
\shorttitle{} 
\shortauthors{} 
 
\begin{document} 
 
\received{} 
\accepted{} 
 

\title{KIC 8262223: A Post-Mass Transfer Eclipsing Binary Consisting of a Delta Scuti Pulsator and a Helium White Dwarf Precursor}  
 
\author{Zhao Guo, Douglas R. Gies, Rachel A. Matson} 
\affil{Center for High Angular Resolution Astronomy and  
 Department of Physics and Astronomy,\\ 
 Georgia State University, P. O. Box 5060, Atlanta, GA 30302-5060, USA; \\guo@chara.gsu.edu, gies@chara.gsu.edu, rmatson@chara.gsu.edu, } 

\author{Antonio Garc\'{i}a Hern\'{a}ndez} 
\affil{Instituto de Astrof{\'i}sica e Ci\^{e}ncias do Espaco, Universidade do Porto, CAUP, Rua das Estrelas, PT4150-762 Porto, Portugal; \\
agh@astro.up.pt} 

\author{Zhanwen Han, Xuefei Chen} 
\affil{Key Laboratory for the Structure and Evolution of Celestial Objects, Yunnan
Observatories,
the Chinese Academy of Sciences, Kunming, 650011, P.R. China \\
zhanwenhan@ynao.ac.cn, cxf@ynao.ac.cn}

\slugcomment{02/17/2017} 

 
\begin{abstract} 
KIC~8262223 is an eclipsing binary with a short orbital period ($P=1.61$ d). The {\it Kepler} light curves are of Algol-type and display deep and partial eclipses, ellipsoidal variations, and pulsations of $\delta$ Scuti type.
We analyzed the {\it Kepler} photometric data, complemented by phase-resolved spectra from the R-C Spectrograph on the 4-meter Mayall telescope  at Kitt Peak National Observatory and determined the fundamental parameters of this system. The low mass and oversized secondary ($M_2=0.20M_{\odot}$, $R_2=1.31R_{\odot}$) is the remnant of the donor star that transferred most of its mass to the gainer, and now the primary star. The current primary star is thus not a normal $\delta$ Scuti star but the result of mass accretion from a lower mass progenitor. We discuss the possible evolutionary history and demonstrate with the MESA evolution code that this system and several other systems discussed in prior literature can be understood as the result of non-conservative binary evolution for the formation of EL CVn type binaries. The pulsations of the primary star can be explained as radial and non-radial pressure modes. The equilibrium models from single star evolutionary tracks can match the observed mass and radius ($M_1=1.94M_{\odot}$, $R_1=1.67R_{\odot}$) but the predicted unstable modes associated with these models differ somewhat from those observed. We discuss the need for better theoretical understanding of such post-mass transfer $\delta$ Scuti pulsators.
\end{abstract} 
 
 
 

\section{Introduction}                                
The $\delta$ Scuti variables are A- or F-type main-sequence (MS) and post-MS stars. They are interesting asteroseismic targets as they are numerous and often show multi-periodic pressure mode pulsations. However, it is hard to determine their fundamental parameters, and their ubiquitous fast rotation makes this task even more difficult. Those eclipsing binaries (EBs) with a component that is a $\delta$ Scuti pulsator offer us the means to determine accurate masses and radii for these variables. However, systems with short orbital periods are more likely to be observed and these are also more likely to experience binary mass transfer. Many $\delta$ Scuti pulsating stars in Algol-type (oEA) systems were discovered from ground-bases photometry (Mkrtichian 2002, 2003; Soydugan et al.\ 2006). The daily gaps in the time series limit the number of detected pulsation frequencies to only a few. Space missions like {\it Kepler} and {\it CoRoT} have provided continuous and accurate light curves of hundreds of $\delta$ Scuti pulsating EBs, and on average, tens of frequencies have been detected for each system. Most of the systems that have been studied in detail were considered as single stars and understood from the theory of single star evolution (Hambleton et al. 2013; Maceroni et al. 2014; da Silva et al. 2014; Guo et al. 2016). In this paper, we focus on $\delta$ Scuti EBs which have undergone mass transfer through the course of binary evolution. KIC~10661783 (Southworth et al.\ 2011; Lehmann et al.\ 2013) belongs to this class, and this binary consists of an oversized low mass secondary ($M=0.2 M_{\odot}, R=1.13R_{\odot}$) and a $\delta$ Scuti pulsating primary.

KIC~8262223 ($K_p=12.146$, $\alpha_{2000}$=$20$:$01$:$19.788$, $\delta_{2000}$=$+44$:$08$:$38.90$) is included in the Kepler Eclipsing Binary Catalog (Pr\v{s}a et al.\ 2011; Slawson et al.\ 2011). It is described as a semi-detached eclipsing binary with an orbital period of $1.612$ days, a near circular orbit, and a high inclination ($\sin i =0.97$). The eclipse timing analysis of this system was performed by Gies et al.\ (2012, 2015) and Conroy et al.\ (2014). The flat $O-C$ diagram of the timings indicates that this circular binary is not likely to have a nearby third companion. Gies et al.\ (2012) also noticed a pulsation signal in near resonance with the orbit in the light curve. Armstrong et al.\ (2014) estimated the effective temperatures of the primary and secondary as
$T_{\rm eff1}=9325\pm 428$ K	
 and $T_{\rm eff2}=6791\pm 642$ K, respectively, based on their fit of the binary spectral energy distribution (SED).

Here we analyze the photometric and spectroscopic data of KIC 8262223 (Section 2, 3, 4), study its pulsations (Section 5), and show that this binary and other similar systems like the aforementioned KIC 10661783 are the products of binary evolution with non-conservative mass-transfer (Section 6). They belong to the class of EL CVn stars which consist of an A- or F-type dwarf primary and a low mass He white dwarf precursor.

\section{ {\it Kepler} Photometry and Ground-based Spectroscopy}  

Simple Aperture Photometry (SAP) data (Data Release 23) from the {\it Kepler} satellite were retrieved from the MAST Archive.
There are 18 quarters (Q0-17) of long cadence data and 1 quarter (Q4) of short cadence data. The long and short cadence data have time sampling rate of $58.8$  seconds and $29.4$ minutes, respectively. Please refer to Caldwell et al.\ (2010) for more information on the {\it Kepler} data.
The aperture contamination factors ($k$) reported in the Kepler Input Catalog are lower than $0.7\%$ in all quarters except for Q12 ($1.3\%$). The light curve shows deep eclipses, ellipsoidal variations and coherent pulsations at frequency about $65$ d$^{-1}$  as apparent in the short cadence data. After removing the long-term trends and outliers (see Section 4), we show a sample of the short cadence light curve in Figure 1. As the main pulsational frequency is well above the Nyquist frequency ($\approx 24$ d$^{-1}$) of long cadence data, these pulsations are essentially all cancelled out in the phase-folded multiple-quarter long cadence data used for binary light curve modeling. The light curve residuals of the short cadence data are used for pulsational analysis.

We also obtained 13 ground-based spectra of moderate resolving power ($R \approx 6000$) from the R-C Spectrograph on Kitt Peak National Observatory (KPNO) 4-meter Mayall telescope from $2010$ to $2013$. The spectra cover the wavelength from $3930$ \AA \ to $4600$ \AA, with typical signal-to-noise ratio of $70-120$ (Figure 2).
More information about the instrument and spectra can be found in Matson et al.\ (2016).

\section{Spectroscopic Orbit and Atmospheric Parameters}

The radial velocities (RVs) were determined following the same cross-correlation technique described by Matson et al.\ (2016). Two templates from atmospheric model grids UVBLUE (Rodr\'{i}guez-Merino et al.\ 2005) were cross-correlated with the observed spectra to obtain the radial velocities presented in Table 1. The derived RVs were fitted to get the orbital parameters ($K_1,K_2,\gamma_1,\gamma_2,T_0$) with the Levenberg-Marquardt algorithm, where $K_1,K_2$ and $\gamma_1,\gamma_2$ are semi-amplitude velocities and system velocities of the primary and secondary star, respectively; $T_0$ is time epoch of the primary minimum (Figure 3). We assumed a circular orbital solution ($e=0$) and the orbital period was fixed to the value from eclipse timing measurements in Gies et al.\ (2015) as $P= 1.61301476$ days. We then used the tomography algorithm (Bagnuolo et al.\ 1994) to reconstruct the individual component spectra of each star. These spectra were compared with a grid of UVBLUE synthetic spectra and the best atmospheric parameters ($T_{\rm eff}$, $\log g$, [Fe/H] and $v\sin i$) were determined from a grid search followed by a local optimization with the Levenberg-Marquardt algorithm. To break the degeneracy in fitting five atmospheric parameters, the $v\sin i$ values were initially estimated from the metal lines in five different spectral sections and the $\log g$ values were fixed to the result from the binary modeling (see next section). Note that the uncertainties were estimated from the covariance matrices, and can be somewhat underestimated. 
The procedures mentioned above are iterative, and in each step the templates and RVs were updated from previous determinations. We adopted the final values when the parameters converged.

Spectral disentangling is another way of deriving orbital parameters. For spectroscopic binaries, we observe the linear combination of two component spectra with different Doppler shifts. Given the radial velocities of the two stars  and their mean flux ratio, we can form a coefficient matrix $A$. Then we can separate the component spectra by solving the linear inverse problem $y=Ax$, where $y$ and $x$ are vectors formed by concatenating the observed composite spectra and the individual component spectra (see Hensberge at al.\ 2008). If the RVs used in the coefficient matrix are calculated from orbital parameters, we can find the optimized orbital parameters by minimizing the $\chi^2$ differences between the observed and synthetic composite spectra, $|y-AX|^2$. We implemented this method with the FDBinary code (Ilijic et al.\ 2004). Note that the code uses a downhill simplex optimizer and regrettably does not provide uncertainty estimates.

The final orbital parameters are summarized in Table 2. The orbital parameters from the two techniques agree very well. The results show that the system has a very small mass ratio ($q=0.104$), and the systemic velocities from fitting RVs of primary and secondary ($\gamma_1,\gamma_2$) agree within uncertainties.
Table 3 contains the optimal atmospheric parameters. This binary consists of a hot A-type primary ($T_{\rm eff1}=9128$ K) and a much cooler secondary ($T_{\rm eff2}=7119$ K). Both stars have metallicities slightly lower than solar. The projected rotational velocity of the primary star ($v\sin i=37\pm 13$ km s$^{-1}$) is  a little lower than the synchronized value at $50$ km s$^{-1}$ (see Table 4 in next section). The $v\sin i$ of the secondary matches the synchronized value very well. Note that each pixel in our spectra is equal to $26.25$ km s$^{-1}$ in velocity space, and we cannot reliably measure small rotational velocities ($v\sin i < 30$ km s$^{-1}$). In Figure 4, we show the reconstructed component spectra of the two stars and the best matching model spectra. The mean flux ratio ($F_2/F_1$) in the observed spectral range ($\approx 4225$ \AA) is $0.21\pm 0.02$ which amounts to percentage contributions of $82.6\%$ and $17.4\%$ for the primary and secondary, respectively.

\section{ELC Binary Models}
We used the {\it Kepler} long cadence data to perform our light curve modeling. The preparation of the raw data, which was detailed in Guo et al. (2016), includes de-trending and outlier removal. We divided the $18$ quarters into eight sections ($Q0-Q2$, $Q3-4$, $Q5-6$, $Q7-8$, $Q9-10$, $Q11-12$, $Q13-14$, $Q15-17$) and fitted the light curve of each individually. The standard deviations of the best fitting parameters from these eight datasets are adopted as the final uncertainties.

We used the Eclipsing Light Curve (ELC) code by Jerome Orosz (Orosz \& Hauschildt 2000) to model the binary light curve. The code implements the Roche model and synthesizes the binary light curve and radial velocity curve by integrating the specific intensity and flux-weighted RVs of each segment on the stellar surface. In ELC, the effect of aperture contamination factor ($k$) is accounted for by adding to the median value of the model light curve $y_{med}$ an offset $ky_{med}/(1- k)$.

We optimized the
following fitting parameters: orbital inclination ($i$), temperature ratio ($ temprat=T_{\rm eff2}/T_{\rm eff1}$), filling factors ($f_1, f_2$) and time of secondary minimum ($T_0$) by implementing the genetic algorithm PIKAIA (Charbonneau 1995). Note that the Roche lobe filling factor ($f$) is defined as $x_{\rm point}/x_{L1}$, where $x_{\rm point}$ is the radius of the star toward the inner Lagrangian point (L1), and $x_{L1}$ is the distance to L1 from the center of the star. It is the counterpart of the Roche potential $\Omega$ used in the Wilson-Devinney code (Wilson \& Devinney 1971). We run the genetic optimizer for 400 generations with 100 members in each generation.  We set broad search ranges for these parameters: $i\in[50,90]$(degrees), $temprat\in[0.6,0.9]$, $f_1,f_2\in[0.1,0.8]$, $T_0\in[55430.9,55432.5]$(BJD-2,400,000). The orbital period was fixed to $1.61301476$ days as found by Gies et al. (2015). The effective temperature of the primary was fixed to the value from spectroscopy ($9128$ K) as it is well known that the light curve from  \textbf{a} single passband is only sensitive to the temperature ratio (if both primary and secondary eclipses occur). 
We assumed the binary has a circular orbit and the two components have synchronized rotation as indicated from spectroscopy. The parameters mass ratio ($q=M_2/M_1$), velocity semi-amplitude ($K_1$), and systemic velocity ($\gamma$) were fixed to values from spectroscopic orbital solutions as they have little affect on the light curve (except for $q$, which can have some influence on the ellipsoidal variations). 
The gravity brightening coefficients ($\beta$) were fixed to the canonical values of $0.25$ for radiative atmospheres and $0.08$ for convective atmospheres.
Similarly, the surface bolometric albedos ($l_1,l_2)$  were set to 1.0 and 0.5 for radiative and convective atmospheres, respectively. 
However, the light curve residuals from the parameter settings above still show obvious variations. We found that by setting $l_2$ as a free parameter the light curve fit is much better. The optimal value of $l_2$ is $0.22$, which is much lower than the canonical value of $0.5$. Note Matson et al.\ (2016) also found a lower albedo ($0.33$) for the F stars in KIC~5738698. If we let the albedo of the primary star ($l_1$) vary, the best value is very close to $1.0$, and the light curve fit is not improved.

Doppler boosting or beaming is a relativistic effect (Loeb \& Gaudi 2003), in which the observer will receive a higher photon rate from a star moving towards him or her, and vice versa. 
The fractional change of photon rate is $\Delta n_{\lambda}/n_{\lambda}=f_{\rm DB}v_{\lambda}/c$, where $v_{\lambda}$ is radial velocity of the star and $c$ is speed of light. Thus, the key parameters are the mass and flux ratios. If the two stars have similar temperatures, then the beaming effect will be canceled out if they have a mass ratio of $1$. For systems with a very small mass ratio, the Doppler beaming effect is expected to play an important role. A measurement of the beaming amplitude from the light curve can provide an independent estimation of the orbital parameters. This was performed in many beaming binaries such as KOI-74, KOI-81 (van Kerkwijk et al.\ 2010) and KIC 11558725 (Telting et al.\ 2012). In the ELC code, the Doppler beaming effect is accounted for following the treatment in van Kerkwijk et al. (2010). The beaming parameter $f_{\rm DB}$ is estimated as the wavelength average of $xe^x/(e^x-1)$ in the {\it Kepler} passband, where $x=hc/(\lambda kT)$. The estimated values are $2.76$ and $3.48$ for the primary and secondary, respectively, and are fixed in the fitting process.
In Figure 5, we show the best light curve solution and corresponding residuals with (red) and without (green) Doppler beaming. It can be seen in the two middle panels that the residuals are more symmetric around the zero horizontal line if the beaming effect is included. We also show the ELC model of the Doppler beaming signal in the bottom panel, and the amplitude of the beaming effect is about $0.00033$ magnitude. 

We also fit the light curves and RVs simultaneously (LC+RV) with fitting parameters $(T_0,i,f_1,f_2,temprat, q,K_1,\gamma)$. Due to the sharp difference between the data quality, we have to give more weight to the RVs. We scaled the errors of light curves so that the $\chi^2_{min} \sim \nu$, where $\nu$ is the degree of freedom. The $1\sigma$ uncertainties were then derived from changing the parameters so that the $\chi^2$ increase by $1.0$ from $\chi^2_{min}$.

In Table 4, we list the final model parameters of KIC~8262223. The parameters of the primary are typical for a mid-A type ZAMS star ($M_1=1.94M_{\odot}$, $R_1=1.67R_{\odot}$, $T_{\rm eff1}=9128$ K) but somewhat over-luminous. The secondary has a very low mass ($M_2=0.20M_{\odot}$) and very discrepant radius ($R_2=1.31R_{\odot}$) and effective temperature ($T_{\rm eff2}=6849$ K) compared to main sequence stars of the same mass. This suggests that this system has gone through binary evolution with mass transfer. The implications and possible evolutionary scenarios are discussed in section 6. The model parameters from fitting light curves and RVs simultaneously (LC+RV) are almost the same as the LC-only values.

The optimal effective temperature ratio ($T_{\rm eff2}/T_{\rm eff1}$) from fitting the light curve is $0.75$, and this gives $T_{\rm eff2}=6849$ K which is $270$ K ($1.8\sigma$) cooler than that from spectroscopy ($T_{\rm eff2}=7119 $ K). This discrepancy can be explained by our adopted lower albedo $l_2=0.22$. There is a correlation between $T_{\rm eff2}$ and bolometric albedo $l_2$. It is known that the bolometric albedo is difficult to pin point and is usually treated as a free parameter. Sometimes even values as high as $2.46$ are used (e.g., star A in KIC~10661783; Lehmann et al. 2013). Note that the effective temperature of KIC~3858884 star B from spectroscopy by Maceroni et al. (2014) is also $\approx 300$ K different from that from the light curve solution. Thus, we think this minor discrepancy is not a problem with our analysis. 
 
\section{Pulsational Characteristics}

We only use the short cadence data to study the pulsations of this system. We calculated the Fourier spectrum of the light curve residuals  with the {\it Period 04} package (Lenz \& Breger 2005) with all eclipses removed. The calculation was performed to the short cadence Nyquist frequency ($\approx 734$ d$^{-1}$) with the fitting formula $Z+\sum_i A_i \sin (2\pi \Omega_i t+2\pi \Phi_i)$, where $Z, A_i,\Omega_i, \Phi_i$ are the zero-point shift, pulsation amplitudes, linear frequencies and phases, respectively, and time $t= $ BJD $ - \ 2,400,000$. No significant peaks were found beyond  $70$ d$^{-1}$. All frequencies with $S/N>4$ are reported in Table 5 and the Fourier spectrum is shown in Figure 6. The uncertainties were calculated following Kallinger et al.\ (2008).

Almost all the pulsations are in the range of $50-65$ d$^{-1}$. There appear to be some low amplitude peaks at $100-130$ d$^{-1}$ (not shown in Figure 6) which are exactly twice the main pulsation range. We interpret these peaks as the harmonics of the main pulsations rather than some high intrinsic pulsation frequencies. This indicates that the pulsations are to some extent non-sinusoidal. The primary star contributes much more light (83\% in the wavelength range of our spectra), and its fundamental parameters ($M_1=1.94M_{\odot}$, $R_1=1.67R_{\odot}$, $T_{\rm eff1}=9128$ K) also agree with those of a typical $\delta$ Scuti pulsator. It is, thus, very likely that the pulsations stem from the primary.  The pulsations at $50-65$ d$^{-1}$ can be well explained as high order ($n_p \approx 6,7$) radial and non-radial p-modes.

In the low frequency region, there are two peaks $f_{20}=1.2397$ d$^{-1}$ and $f_{58}=0.79928$ d$^{-1}$. $f_{20} $ is equal to twice the orbital frequency $2f_{\rm orb}=2\times 0.61996$ d$^{-1}$ within uncertainties and is likely the result of imperfect light curve fitting (e.g., ellipsoidal variations). $f_{58}$ is probably an artifact of imperfect data reduction.

As the star is pulsating at relatively high radial orders, which are closer to the asymptotic regime, we can expect to find some frequency regularities similar to those observed in solar-like oscillators.
Garc{\'{\i}}a Hern{\'a}ndez et al.\ (2015) found the signature of frequency regularities in six $\delta$ Scuti stars in eclipsing binaries by analyzing the Fourier Transform (FT) of the p-mode frequencies. These frequency patterns are close to the large frequency separation which is related to the mean stellar density. We applied the same FT technique to the frequencies in Table 5, and the Fourier spectrum is presented in Figure 7. The periodicity at $39.9\mu \rm$HZ (3.45 d$^{-1}$) is related to half of the large frequency separation $0.5 \Delta \nu$. From the mean density of the primary, we can deduce the expected large separation ($\Delta \nu$) by using the linear relation between $\log \Delta \nu$ and $\log \rho$ (Su{\'a}rez et al.\ 2014; Garc{\'{\i}}a Hern{\'a}ndez et al.\ 2015). The expected $\Delta \nu$ is $70.7$ $\mu$HZ (6.1 d$^{-1}$), which is similar to but smaller than the observed value $\Delta \nu_{\rm obs}=2\times 39.9=79.8$ $\mu$HZ (6.89 d$^{-1}$). Paparo et al.\ (2016a, b) found the signatures of the large frequency separation in $90$ $\delta$ Scuti stars observed by {\it CoRoT} satellite, and in addition to showing regularities of $\Delta\nu$, some of them show patterns which approximately agree with $\Delta\nu\pm f_{\rm rot}$ or $\Delta\nu\pm2 f_{\rm rot}$. If we adopt the synchronous rotational frequency $f_{\rm rot}=7.175\mu$HZ (0.62 d$^{-1}$), the corresponding rotational splittings for the high order p-modes are $m(1-C_{nl}) f_{\rm rot}\approx m f_{\rm rot}$ (Aerts et al.\ 2010), and $m=\pm1$ and $m=\pm1, \pm2$ for $l=1$ and $l=2$ modes, respectively. We thus conclude that the observed pattern $\Delta\nu_{\rm obs}$ agrees with the theoretical large frequency separation with rotational effect taken into account. Indeed, the highest peak in the Fourier spectrum in Figure 7 is at about $7.07\mu$HZ (0.61 d$^{-1}$) which is only slightly smaller than the orbital frequency $f_{\rm orb}=7.175\mu$HZ (0.62 d$^{-1}$), and thus this regularity is likely the result of rotational splitting. 

Tidally excited g-mode oscillations have been found in many eccentric binaries (Welsh et al.\ 2011; Hambleton et al.\ 2016; Guo et al.\ 2017). For KIC 8262223 with a synchronized circular orbit, this effect of the dynamical tide is not expected. Note that tidal oscillations can also enforce frequency splitting of p-modes at integer multiples of orbital frequency (e.g., the eccentric binary KIC4544587 in Hambleton et al.\ 2013) through mode coupling of self-excited p-modes and tidally induced g-modes.

To check whether the pulsation range can be explained by the non-adiabatic theory, we modeled the evolution of some single non-rotating stars with MESA (Paxton et al. 
2011, 2013) and calculated their pulsation frequencies in the range of $20-70$ d$^{-1}$ 
with GYRE (Townsend \& Teitler 2013). We set the mixing length parameter $\alpha_{\rm MLT}$ to $1.8$ and used the OPAL opacity tables (Iglesias \& Rogers 1996). Solar 
mixtures in Grevesse \& Sauval (1998) were adopted for the assumed solar composition. The results are presented in Figure 8. The pulsation modes ($l=0, 1, 2$) from the equilibrium model with $M=1.94M_{\odot}$, $R=1.67M_{\odot}$, $Y=0.28$, and $Z=0.02$ are unstable ($\eta > 0$, the normalized growth-rate defined in Stellingwerf 1978) in the range of $60-67$ d$^{-1}$, 
which agrees approximately with the observed unstable range $50-65$ d$^{-1}$. However, the theoretical unstable range is much narrower than observations. We have to be cautious in interpreting the above analysis based on single star evolution, because the real inner structure of the $\delta$ Scuti type primary may have different pulsational properties due to interior changes caused by the past mass transfer in the binary. Our preliminary analysis of these post-mass transfer $\delta$ Scuti stars suggests that they tend to be hotter and pulsate over a broader range and with higher frequencies than their single-evolution counterparts with the same mass and radius. Thus the rejuvenation of $\delta$ Scuti stars from binary evolution is a possible candidate to explain the overabundance of high-frequency $\delta$ Scuti pulsators (Balona et al.\ 2015). This will be presented in a separate paper (Z. Guo in preparation).

\section{Evolution}

The primary star of KIC~8262223 appears \textbf{to be} a normal A-type dwarf near the zero age main sequence (ZAMS) (slightly over-luminous), while the low-mass secondary is noticeably oversized and over-luminous. The classical scenario for the formation of this type of cool Algol system involves the mass transfer (probably case B, i.e., the donor star fills its Roche lobe while evolving through the Hertzsprung gap, Paczynski 1971) from the original massive primary (donor) to the original less massive secondary, leading to a mass ratio reversal.

KIC~8262223 is likely to evolve into a typical EL CVn system, which consists of a normal A- or F-type dwarf and a low mass ($\approx 0.2 M_{\odot}$) helium white dwarf precursor (pre-He WD). Maxted et al.\ (2014) presented 17 EL CVn systems discovered by the WASP survey. 
KIC~8262223 closely resembles the cool Algol system KIC~10661783 described by Southworth et al.\ (2011) and Lehmann et al.\ (2013). The latter authors also discussed several similar systems such as AS Eri (Mkrtichian et al.\ 2004) and V228 (Kaluzny et al.\ 2007). Sarna et al.\ (2008) found that a system with similar initial masses and slightly longer period ($M_{10}=0.88M_{\odot},M_{20}=0.85M_{\odot}, P=1.35$d) can evolve to the current state of V228 ($M_{2}=0.20M_{\odot},M_{1}=1.51M_{\odot}, P=1.15$d) through non-conservative case B mass transfer (see also Stepien et al.\ 2016). Eggleton \& Kiseleva-Eggleton (2002) studied binary evolution of cool Algols including AS Eri. For better comparison, we list the parameters of four systems KIC 8262223, KIC 10661783, AS Eri, and V228 in Table 6. All these binaries consist of a low mass secondary ($\approx 0.2 M_{\odot}$) and may have similar evolutionary history as detailed below. 

In Figure 9, we show the positions of the above four systems and the evolutionary tracks of He-WDs calculated by Driebe et al.\ (1999). The $\log g$ and $T_{\rm eff}$ of KIC 8262223 are fitted nicely by the evolutionary track of mass $0.195M_{\odot}$, matching the observed mass from RVs ($0.20 \pm 0.01 M_{\odot}$). The observed quantities of other three systems also agree with the theory if the uncertainties of masses are considered.

Chen et al.\ (2016) found that EL CVn type binaries can result from non-conservative binary evolution with long-term stable mass transfer
between low-mass stars \textbf{that avoided} a rapid common-envelope evolution. They did thorough simulations with the MESA code  (Paxton et al.\ 2011, 2013) to analyze the evolution channel of EL CVn stars from low mass progenitors ($M_{10}\in[0.9,2.0]M_{\odot}$ and $q_0=M_{10}/M_{20} \in [1.1,4.0]$). The parameters of the secondary star of KIC~8262223 ($R_2/a=0.176$, $T_{\rm eff2}=6849K$) fit their $R_2/a-T_{\rm eff2}$ relation for pre-He WD very well (see their Fig. $9$). They also found a tight correlation between orbital periods and WD masses as shown in their Figure $10$. Our observed values ($P=1.6$d, $M_2=0.20M_{\odot}$) also nicely match their theoretical relations. According to their Figure $7$, the pre-He WD in KIC~8262223 with a mass of $0.2M_{\odot}$ has an envelope mass of $0.02M_{\odot}$.


Due to the uncertainties in the treatment of mass loss and angular momentum loss of binary evolution, we do not attempt to find a best matching evolution history for KIC~8262223. Instead, we show that this binary as well as other binaries in Table 6 can be qualitatively explained by the aforementioned formation channel.
We used the binary module of MESA evolution code (v7624) and evolved two typical systems: (1) $M_{10}=1.35M_{\odot}$, $M_{20}=1.15M_{\odot}$ and $P_0=2.89$ d; (2) $M_{10}=1.0M_{\odot}$, $M_{20}=0.9M_{\odot}$ and $P_0=3.0$d. The metallicities were set to the solar value ($Z=0.02$) and initial helium abundances were fixed to $Y=0.28$. The evolutionary tracks were assumed to be non-conservative. Following the assumptions in Chen et al.\ (2016), half of the mass
lost from the vicinity of the donor is accreted by the gainer while the other 50\% leaves the system as a fast wind, carrying away the same angular momentum \textbf{as} the donor. The mass transfer rate is calculated implicitly using the Ritter scheme (Ritter 1988).

The evolutionary paths are shown in the H-R diagram in Figure 10. The black and red tracks are for the donor and gainer in model (1), respectively. 
The system starts with an orbital period of $2.89$ d, and the two stars follow their single star evolutionary tracks. The mass transfer begins when the primary evolves to the sub-giant stage and its radius reaches its Roche lobe (at $t=3.32$ Gyr, marked by the filled circles). After a stable mass-transfer of about $\sim 1$ Gyr (marked by the star symbol), the donor star evolves to a stable long-term stage of almost constant luminosity and begins to contract, cool, and evolve to a He WD precursor. The system ends up with parameter values $M_{1}=0.218M_{\odot},M_2=1.716M_{\odot}$, and $P=3.59$ d. For model (2), the evolution of the initial primary star $M_1$ (gray line, secondary evolution not shown for clarity) is shown in Figure 10. The final status of this system has parameters of
$M_{1}=0.20M_{\odot},M_2=1.30M_{\odot},P=1.06$ d. 

The observed positions of four cool Algol systems mentioned above are shown in Figure 10, with mass-gainers indicated as red symbols and donors as black symbols. According to the above evolutionary models,  the mass gainer evolves along the red track to the upper left and arrives at the observed locations of the A- or F-type dwarfs which can enter the $\delta$ Scuti instability strip (IS). The blue and red edges for the fundamental radial modes calculated by Dupret et al.\ (2005) are denoted by the dotted lines, and those for the fourth overtone radial modes are marked by the dashed lines\footnote{The blue/red edges depend on the radial order of the modes, as well as other model parameters (e.g., mixing length parameter $\alpha_{\rm MLT}$) KIC 8262223 seems to reside outside the instability strip for $n_p=4$, and it may pulsate at higher order ($n_p \approx 7-8$). }. The mass donor gets hotter, smaller, and evolves to the left and becomes a He WD precursor. The evolutionary tracks of these two representative models are not meant to explain quantitatively the properties of the four systems, but rather to show the regions that the product of the binary evolution can occupy on the H-R diagram and the final status for the formation of EL CVn stars. For KIC~8262223, the secondary seems to be a star that has just finished its mass transfer and is contracting (i.e., from filling its Roche lobe to 
under-filling its Roche lobe). The secondary still has a large radius, and thus the binary light curves show a partical eclipse instead of a flat-bottomed transit signal which is typical for EL CVn stars.

It is interesting to note that the dwarf stars in these systems are often pulsating (all but V228, which has a mass too low to be a $\delta$ Scuti pulsator). As can be seen in Table 6, these systems can pulsate at low ($20-30$ d$^{-1}$) as well as high frequencies ($50-60$ d$^{-1}$). It is known that the unstable range of pulsations will vary as the star evolves off the ZAMS. For example, for a $\delta$ Scuti star with $M=1.8M_{\odot}$, the $p$-modes $n_p=4-7$ ($45-60$ d$^{-1}$) are unstable for young models close to ZAMS. The unstable range moves to $5-25$ d$^{-1}$ for models near TAMS which are low order p-modes or g-modes (Dupret 2002). Asteroseismology has the potential of determining the ages of the $\delta$ Scuti pulsators in these EL CVn binaries. Not only the dwarfs, but the pre-He WD precursor can also show pulsations. One such example is the $g$-mode pulsating WD in KIC 9164561 (Zhang et al.\ 2016). Such pulsations enabled the discovery of a thick hydrogen envelope on the pre-He WD J0247-25B (Maxted et al.\ 2013). The theoretical instability strip of these pre-He WDs has been examined closely by C{\'o}rsico \& Althaus (2016). It is interesting to investigate whether the mass-gainer $\delta$ Scuti star and the He WD can both reside in their corresponding instabiity strip, as the current observations only reveal systems containing one pulsator. More information can be extracted from these pulsations, which may lead to great advancements in our understanding of the evolution of low mass close binaries.

\section{Conclusions and Prospects}
Utilizing the accurate {\it Kepler} photometric data and our ground-based spectroscopic data, we determined the fundamental parameters of KIC~8262223, an eclipsing binary system with an orbital period of $1.6$ days which contains an A-type dwarf and a low mass pre-He WD. The light curves show high frequency pulsations at about $60$ d$^{-1}$. These $\delta$ Scuti type pulsations are likely from the primary star and can be explained as radial and non-radial p-modes.
We discussed possible evolutionary scenarios and showed that this system and several other very similar binaries can be explained by the non-conservative evolution of close binaries with low mass progenitors, the channel that forms EL CVn type stars. KIC~8262223 also poses some challenges to our non-adiabatic theory of stellar pulsations in modeling these post-mass transfer $\delta$ Scuti stars. These rejuvenated $\delta$ Scuti stars pulsate with high frequencies and may explain the observed over-abundance of high-frequency $\delta$ Scuti stars in the {\it Kepler} field (Balona et al.\ 2015). 

Asteroseismic modeling has not yet been applied to post-mass transfer $\delta$ Scuti stars in pulsating Algols (oEAs) due to the complex nature of these systems. As a prerequisite, it is possible to identify the pulsation modes through high cadence and high resolution spectroscopy. The eclipse mapping method (Reed et al.\ 2005; B{\'{\i}}r{\'o} \& Nuspl 2011) is also promising but still awaits application to a real object. Several hundreds of $\delta$ Scuti variables in binaries have already been detected by the {\it Kepler} satellite as well as ground-based observations (Pigulski \& Michalska 2007), and future missions like {\it TESS} will provide more systems. A complete analysis of their pulsational properties will require a better understanding of close binary tidal interactions and binary evolution.
 
\acknowledgments 
We are in debt to the anonymous referee for helpful comments and suggestions which greatly improved the quality of this paper. We thank Jerome A. Orosz for his constant support in using the ELC code. We thank Bill Paxton, Rich Townsend and others for maintaining and updating MESA and GYRE.
We thank Meng Sun for helpful discussions. This work is partly based on data from the {\it Kepler} mission. Funding for this mission is provided by NASA's Science Mission Directorate.
The photometric data were obtained from the Mikulski Archive for Space Telescopes (MAST). STScI is operated by the Association of Universities for Research in Astronomy, Inc., under NASA contract NAS5-26555. This study was supported by NASA grants NNX12AC81G, NNX13AC21G, and NNX13AC20G. This material is based upon work supported by the National Science Foundation under Grant No. AST-1411654. A. G. H. acknowledges support from Funda\c{c}\~{a}o para a Ci\^{e}ncia
e a Tecnologia (FCT, Portugal) through the fellowship SFRH/
BPD/80619/2011, and
from the EC Project SPACEINN (FP7-SPACE-2012-312844). Z. H. is partly supported by the Natural Science Foundation of China
 (Grant Nos 11521303, 11390374). Institutional support has been provided from the GSU College 
of Arts and Sciences and the Research Program Enhancement 
fund of the Board of Regents of the University System of Georgia, 
administered through the GSU Office of the Vice President 
for Research and Economic Development.

 


\clearpage


\begin{deluxetable}{cccccccc}
\tabletypesize{\small} 
\tablewidth{0pc} 
\tablenum{1} 
\tablecaption{Radial Velocities\label{tab1}} 
\tablehead{ 
\colhead{Time}          & 
\colhead{Phase}&
\colhead{$V_{r}$(pri)}        & 
\colhead{$O-C$}        & 
\colhead{$V_{r}$(sec)}    &
\colhead{$O-C$}        & 
\colhead{Observation}                          \\  
\colhead{(HJD-2400000)}           & 
\colhead{}                 &
\colhead{(km s$^{-1}$)} & 
\colhead{(km s$^{-1}$)}& 
\colhead{(km s$^{-1}$)} & 
\colhead{(km s$^{-1}$)}& 
\colhead{Source}& 
       
} 

\startdata		
       55369.9232 &       0.19 & $       1.5$ $\pm$        1.7 &      $-0.9$&        215.4  $\pm$        4.3&       $0.4$ & KPNO   \\
       55732.8574 &       0.19 & $       2.6$ $\pm$        1.7 &      0.4&        213.5  $\pm$        4.5&      $-3.3$ & KPNO   \\
       55815.8979 &       0.68 & $       42.7$ $\pm$        1.9 &      0.2&       $-159.2$  $\pm$        5.9&      $-2.1$ & KPNO   \\
       56077.9534 &       0.14 & $       6.2$ $\pm$        1.7 &       0.8&        185.0  $\pm$        4.7&      4.2 & KPNO   \\
       56078.7629 &       0.64 & $       42.3$ $\pm$        2.4 &       2.0&       $-144.2$  $\pm$        7.0&      $-11.9$ & KPNO   \\
       56078.8440 &       0.79 & $       42.0$ $\pm$        1.8 &      $-1.2$&       $-163.3$  $\pm$        5.7&       1.9 & KPNO   \\
       56078.9357 &       0.75 & $       45.8$ $\pm$        1.9 &       1.6&       $-183.5$  $\pm$        5.6&      $-3.9$ & KPNO   \\
       56079.7925 &       0.28 & $       9.3$ $\pm$        3.4 &       7.2&        220.4  $\pm$        10.7&       $-6.4$ & KPNO   \\
       56081.9642 &       0.63 & $       34.8$ $\pm$        2.9 &      $-4.2$&       $-103.3$  $\pm$        6.4&       15.5 & KPNO   \\
       56082.8204 &       0.16 & $       3.1$ $\pm$        1.8 &     $-1.0$&        195.2  $\pm$        4.7&      0.9 & KPNO   \\
       56082.8833 &       0.20 & $       1.7$ $\pm$        1.7 &    $-0.4$&        217.2  $\pm$        4.7&      0.0 & KPNO   \\
       56082.9468 &       0.23 & $      0.6$ $\pm$        2.0 &      -0.8&        227.5  $\pm$        5.6&      $-1.2$ & KPNO   \\ 
\enddata 
\end{deluxetable}

\begin{deluxetable}{lccc} 
\tabletypesize{\small} 
\tablewidth{0pc} 
\tablenum{2} 
\tablecaption{Orbital Parameters\label{tab2}} 
\tablehead{ 
\colhead{Parameter}   & 
\colhead{RVs}      &
\colhead{Spectral Disentangling}      &

}
\startdata 
$T_0$ (primary minimum) (HJD-2,400,000)               & $55690.5 \pm 0.1$ & $55690.605$  \\ 
$K_1$ (km s$^{-1}$)               & $21.4 \pm 1.0$ & $21.5$  \\ 
$K_2$ (km s$^{-1}$)    & $204.8\pm 3.2$  & $201.4$       \\ 

$\gamma_1$  (km s$^{-1}$)      & $22.8 \pm 0.6 $        & $...$           \\ 

$\gamma_2$  (km s$^{-1}$)   & $25.1\pm 1.7$ &  $...$        \\ 

$e$  & $0.0\tablenotemark{a}$        & $ 0.0\tablenotemark{a}$        \\ 
\hline
$rms_1$ (km s$^{-1}$)   & $2.6$        & $...$             \\ 

$rms_2$ (km s$^{-1}$)  & $6.3$        & $ ...$             \\ 

\enddata 
\tablenotetext{a}{Fixed.}
\end{deluxetable}

\begin{deluxetable}{lccc} 
\tabletypesize{\small} 
\tablewidth{0pc} 
\tablenum{3} 
\tablecaption{Atmospheric Parameters\label{tab3}} 
\tablehead{ 
\colhead{Parameter}   & 
\colhead{Primary}      &
\colhead{Secondary}      &

}
\startdata 
$T_{\rm eff}$ (K)               \dotfill & $9128 \pm 130$ & $7119 \pm 150$ \\ 

$\log g$ (cgs)  \dotfill & $4.3\tablenotemark{a}$  & $3.5\tablenotemark{a}$       \\ 

$v \sin i$ (km s$^{-1}$)     \dotfill & $37 \pm 13$        & $35 \pm 10$          \\ 

$\rm[Fe/H]$ \dotfill & $-0.05 \pm 0.10$        & $ -0.05 \pm 0.10$                 \\ 
\hline
Flux Contribution \dotfill & $82.6\%$     &$17.4\%$   
\enddata 
\tablenotetext{a}{Fixed.}

\end{deluxetable} 

\begin{deluxetable}{lcccc} 
\tabletypesize{\small} 
\tablewidth{0pc} 
\tablenum{4} 
\tablecaption{Model Parameters\label{tab3}} 
\tablehead{ 
\colhead{Parameter}   & 
\colhead{Solution (LC) }    &
\colhead{Solution (LC+RV) }          
}
\startdata 
Period (days) & $1.61301476\tablenotemark{a}$& $1.61301476\tablenotemark{a}$\\
Time of primary minimum (BJD-2400000) & $55432.522844(7)$ &$55432.522844(7)$   \\
Mass ratio $q=M_2/M_1$               &$0.104\tablenotemark{a}$ &0.107(2)    \\ 
Orbital eccentricity $e$              &$0.0\tablenotemark{a}$  &$0.0\tablenotemark{a}$      \\
Orbital inclination $i$ (degree) &$75.203(7)$ & 75.178(2)\\
Semi-major axis a ($R_\odot$) &$7.45(11)$ &7.48(10)\\
$M_1$ ($M_\odot$)              & $1.94(6)$ &1.96(6)               \\ 
$M_2$ ($M_\odot$)              & $0.20(1)$ &0.21(1)             \\ 
$R_1$ ($R_\odot$)               & $1.67(3)$   &1.67(4)        \\
$R_2$ ($R_\odot$)               & $1.31(2)$  &1.32(3)        \\
Filling factor $f_1$     &$0.314(3)$ &$0.314(2)$    \\
Filling factor $f_2$     &$0.672(1)$  &0.671(1)  \\
Gravity brightening, $\beta_1$     &$0.25\tablenotemark{a}$ &$0.25\tablenotemark{a}$    \\
Gravity brightening, $\beta_2$     &$0.08\tablenotemark{a}$&$0.08\tablenotemark{a}$   \\
Bolometric albedo 1& $1.0\tablenotemark{a}$&$1.0\tablenotemark{a}$       \\

Bolometric albedo 2& $0.22(1)$   &0.22(1)    \\

Beaming parameter 1& $2.76\tablenotemark{a}$&$2.76\tablenotemark{a}$     \\

Beaming parameter 2& $3.48\tablenotemark{a}$&$3.48\tablenotemark{a}$      \\

$T_{\rm eff1}$ (K)                    & $9128\tablenotemark{a} $&$9128\tablenotemark{a} $ \\

$T_{\rm eff2}$ (K)                    & $6849(15) $ &$6885(24) $ \\

$\log g_1$ (cgs)  & $4.28(4)$ & $4.28(2)$         \\ 
$\log g_2$ (cgs)  & $3.51(6)$ & $3.52(2)$        \\ 

Synchronous $v \sin i_1$ (km s$^{-1}$)   &   $50.6(9)$ &50.7(9)  
           \\ 
Synchronous $v \sin i_2$ (km s$^{-1}$)   &   $39.6(6)$  &39.9(7) 
           \\ 
$K_1$ (km s$^{-1}$)   &   $21.4\tablenotemark{a}$  &21.9(3) 
           \\ 
$\gamma$ (km s$^{-1}$)   &   $22.8\tablenotemark{a}$  &23.0(5) 
           \\ 
\enddata 
\tablenotetext{a}{Fixed.}
\end{deluxetable}

\begin{deluxetable}{lccccccc} 
\tabletypesize{\small} 
\tablewidth{0pc} 
\tablenum{5} 
\tablecaption{Significant oscillation frequencies\label{tab5}} 
\tablehead{ 
\colhead{}   & 
\colhead{Frequency (d$^{-1}$)}      &
\colhead{Amplitude ($10^{-3}$)}      &
\colhead{Phase (rad/$2\pi$)} & 
\colhead{S/N} &
\colhead{Comment} &
}
\startdata 
$f_{       1}$    & $64.43390 \pm 0.00010$ & $1.319 \pm 0.020$ & $0.896\pm 0.007$ & $114.5$ & $$ \\
$f_{       2}$    & $57.17794 \pm 0.00016$ & $0.918 \pm 0.022$ & $0.313\pm 0.011$ & $72.3$ & $$ \\
$f_{       3}$    & $61.43616 \pm 0.00018$ & $0.782 \pm 0.021$ & $0.190\pm 0.012$ & $64.4$ & $$ \\
$f_{       4}$    & $53.64792 \pm 0.00024$ & $0.620 \pm 0.022$ & $0.345\pm 0.016$ & $49.0$ & $$ \\
$f_{       5}$    & $51.04548 \pm 0.00026$ & $0.565 \pm 0.021$ & $0.281\pm 0.017$ & $46.0$ & $$ \\
$f_{       6}$    & $54.78183 \pm 0.00028$ & $0.540 \pm 0.022$ & $0.368\pm 0.019$ & $42.5$ & $$ \\
$f_{       7}$    & $63.28439 \pm 0.00028$ & $0.497 \pm 0.020$ & $0.516\pm 0.019$ & $42.2$ & $$ \\
$f_{       8}$    & $60.31265 \pm 0.00040$ & $0.366 \pm 0.021$ & $0.949\pm 0.027$ & $29.7$ & $$ \\
$f_{       9}$    & $61.19863 \pm 0.00040$ & $0.363 \pm 0.021$ & $0.896\pm 0.027$ & $29.8$ & $$ \\
$f_{      10}$    & $49.08047 \pm 0.00039$ & $0.357 \pm 0.020$ & $0.218\pm 0.027$ & $30.1$ & $$ \\
$f_{      11}$    & $60.19302 \pm 0.00052$ & $0.284 \pm 0.021$ & $0.669\pm 0.035$ & $22.9$ & $$ \\
$f_{      12}$    & $63.82187 \pm 0.00049$ & $0.281 \pm 0.020$ & $0.947\pm 0.033$ & $24.1$ & $$ \\
$f_{      13}$    & $54.88585 \pm 0.00059$ & $0.255 \pm 0.022$ & $0.607\pm 0.040$ & $20.0$ & $$ \\
$f_{      14}$    & $62.43309 \pm 0.00063$ & $0.226 \pm 0.020$ & $0.717\pm 0.042$ & $18.9$ & $$ \\
$f_{      15}$    & $53.54042 \pm 0.00071$ & $0.212 \pm 0.022$ & $0.134\pm 0.048$ & $16.7$ & $$ \\
$f_{      16}$    & $57.77610 \pm 0.00071$ & $0.210 \pm 0.022$ & $0.885\pm 0.048$ & $16.6$ & $$ \\
$f_{      17}$    & $50.32422 \pm 0.00071$ & $0.201 \pm 0.021$ & $0.513\pm 0.048$ & $16.6$ & $$ \\
$f_{      18}$    & $55.94001 \pm 0.00078$ & $0.193 \pm 0.022$ & $0.050\pm 0.053$ & $15.2$ & $$ \\
$f_{      19}$    & $50.97786 \pm 0.00079$ & $0.185 \pm 0.021$ & $0.755\pm 0.053$ & $15.1$ & $$ \\
$f_{      21}$    & $64.47378 \pm 0.00079$ & $0.172 \pm 0.020$ & $0.584\pm 0.053$ & $14.9$ & $$ \\
$f_{      22}$    & $61.55406 \pm 0.00086$ & $0.167 \pm 0.021$ & $0.746\pm 0.058$ & $13.7$ & $$ \\
$f_{      23}$    & $61.46043 \pm 0.00087$ & $0.165 \pm 0.021$ & $0.255\pm 0.059$ & $13.6$ & $$ \\
$f_{      24}$    & $58.42108 \pm 0.00097$ & $0.153 \pm 0.022$ & $0.494\pm 0.066$ & $12.2$ & $$ \\
$f_{      25}$    & $63.66929 \pm 0.00096$ & $0.144 \pm 0.020$ & $0.252\pm 0.065$ & $12.3$ & $$ \\
$f_{      26}$    & $49.85027 \pm 0.00100$ & $0.143 \pm 0.021$ & $0.713\pm 0.067$ & $11.9$ & $$ \\
$f_{      27}$    & $58.42454 \pm 0.00108$ & $0.138 \pm 0.022$ & $0.593\pm 0.073$ & $10.9$ & $$ \\
$f_{      28}$    & $60.23810 \pm 0.00108$ & $0.135 \pm 0.021$ & $0.568\pm 0.073$ & $10.9$ & $$ \\
$f_{      29}$    & $59.09553 \pm 0.00117$ & $0.127 \pm 0.021$ & $0.757\pm 0.079$ & $10.1$ & $$ \\
$f_{      30}$    & $56.62486 \pm 0.00120$ & $0.126 \pm 0.022$ & $0.992\pm 0.081$ & $9.9$ & $$ \\
$f_{      31}$    & $60.28838 \pm 0.00126$ & $0.116 \pm 0.021$ & $0.751\pm 0.085$ & $9.4$ & $$ \\
$f_{      32}$    & $64.52926 \pm 0.00119$ & $0.114 \pm 0.020$ & $0.752\pm 0.080$ & $9.9$ & $$ \\
$f_{      33}$    & $52.27648 \pm 0.00130$ & $0.114 \pm 0.021$ & $0.204\pm 0.088$ & $9.1$ & $$ \\
$f_{      34}$    & $50.42825 \pm 0.00130$ & $0.111 \pm 0.021$ & $0.808\pm 0.088$ & $9.1$ & $$ \\
$f_{      35}$    & $54.50442 \pm 0.00142$ & $0.105 \pm 0.022$ & $0.515\pm 0.096$ & $8.3$ & $$ \\
$f_{      36}$    & $63.20290 \pm 0.00146$ & $0.096 \pm 0.020$ & $0.717\pm 0.099$ & $8.1$ & $$ \\
$f_{      37}$    & $56.85719 \pm 0.00165$ & $0.091 \pm 0.022$ & $0.197\pm 0.111$ & $7.2$ & $$ \\
$f_{      38}$    & $59.76651 \pm 0.00174$ & $0.084 \pm 0.021$ & $0.690\pm 0.118$ & $6.8$ & $$ \\
$f_{      40}$    & $57.75009 \pm 0.00199$ & $0.075 \pm 0.022$ & $0.430\pm 0.134$ & $5.9$ & $$ \\
$f_{      41}$    & $61.62167 \pm 0.00207$ & $0.069 \pm 0.021$ & $0.177\pm 0.139$ & $5.7$ & $$ \\
$f_{      42}$    & $60.36814 \pm 0.00215$ & $0.068 \pm 0.021$ & $0.742\pm 0.145$ & $5.5$ & $$ \\
$f_{      43}$    & $51.07405 \pm 0.00215$ & $0.068 \pm 0.021$ & $0.854\pm 0.145$ & $5.5$ & $$ \\
$f_{      44}$    & $64.40949 \pm 0.00205$ & $0.067 \pm 0.020$ & $0.325\pm 0.138$ & $5.8$ & $$ \\
$f_{      45}$    & $61.36160 \pm 0.00222$ & $0.065 \pm 0.021$ & $0.574\pm 0.150$ & $5.3$ & $$ \\
$f_{      46}$    & $65.66649 \pm 0.00212$ & $0.063 \pm 0.019$ & $0.513\pm 0.143$ & $5.6$ & $$ \\
$f_{      47}$    & $66.96511 \pm 0.00209$ & $0.062 \pm 0.019$ & $0.258\pm 0.141$ & $5.7$ & $$ \\
$f_{      48}$    & $54.37597 \pm 0.00243$ & $0.062 \pm 0.022$ & $0.028\pm 0.164$ & $4.9$ & $$ \\
$f_{      49}$    & $51.11913 \pm 0.00237$ & $0.061 \pm 0.021$ & $0.195\pm 0.160$ & $5.0$ & $$ \\
$f_{      50}$    & $60.16181 \pm 0.00242$ & $0.060 \pm 0.021$ & $0.567\pm 0.163$ & $4.9$ & $$ \\
$f_{      51}$    & $60.34386 \pm 0.00242$ & $0.060 \pm 0.021$ & $0.157\pm 0.163$ & $4.9$ & $$ \\
$f_{      52}$    & $58.97922 \pm 0.00247$ & $0.060 \pm 0.021$ & $0.686\pm 0.167$ & $4.8$ & $$ \\
$f_{      54}$    & $55.81344 \pm 0.00262$ & $0.057 \pm 0.022$ & $0.227\pm 0.177$ & $4.5$ & $$ \\
$f_{      55}$    & $58.14367 \pm 0.00266$ & $0.056 \pm 0.022$ & $0.960\pm 0.179$ & $4.5$ & $$ \\
$f_{      56}$    & $59.80451 \pm 0.00275$ & $0.054 \pm 0.021$ & $0.532\pm 0.185$ & $4.3$ & $$ \\
$f_{      58}$    & $0.79928 \pm 0.00277$ & $0.052 \pm 0.021$ & $0.658\pm 0.187$ & $4.3$ & $$ \\
$f_{      59}$    & $54.28582 \pm 0.00294$ & $0.051 \pm 0.022$ & $0.767\pm 0.199$ & $4.0$ & $$ \\
$f_{      60}$    & $48.03643 \pm 0.00270$ & $0.051 \pm 0.020$ & $0.140\pm 0.182$ & $4.4$ & $$ \\
$f_{      61}$    & $59.63633 \pm 0.00291$ & $0.051 \pm 0.021$ & $0.109\pm 0.197$ & $4.1$ & $$ \\
$f_{      62}$    & $62.25437 \pm 0.00288$ & $0.049 \pm 0.021$ & $0.425\pm 0.194$ & $4.1$ & $$ \\
$f_{      63}$    & $64.90535 \pm 0.00281$ & $0.048 \pm 0.020$ & $0.340\pm 0.190$ & $4.2$ & $$ \\
$f_{      64}$    & $65.72024 \pm 0.00282$ & $0.047 \pm 0.019$ & $0.585\pm 0.190$ & $4.2$ & $$ \\
\hline
$f_{      20}$    & $1.23967 \pm 0.00079$ & $0.183 \pm 0.021$ & $0.027\pm 0.053$ & $14.9$ & $2f_{orb}$ \\
$f_{      39}$    & $58.09686 \pm 0.00194$ & $0.077 \pm 0.022$ & $0.993\pm 0.131$ & $6.1$ & $f_{37}+2f_{orb}$ \\
$f_{      53}$    & $49.18796 \pm 0.00242$ & $0.058 \pm 0.020$ & $0.677\pm 0.163$ & $4.9$ & $f_{34}-2f_{orb}$ \\
$f_{      57}$    & $62.58379 \pm 0.00267$ & $0.053 \pm 0.020$ & $0.528\pm 0.180$ & $4.4$ & $f_{36}-f_{orb}$ \\

\enddata 
\end{deluxetable} 

\begin{deluxetable}{ccccccccccc}
\tabletypesize{\small} 
\tablewidth{0pc} 
\tablenum{6} 
\tablecaption{Comparison of four cool Algols\label{tab6}} 
\tablehead{ 
\colhead{Name}          & 
\colhead{$M_1$}&
\colhead{$M_2$}        & 
\colhead{$R_1$}        & 
\colhead{$R_2$}    &
\colhead{$T_{\rm eff1}$}        & 
\colhead{$T_{\rm eff2}$}        & 
\colhead{Period}        & 
\colhead{Pulsation}        &     
\colhead{Remark}        &              \\  
\colhead{}           & 
\colhead{($M_{\odot}$)}                 &
\colhead{($M_{\odot}$)} & 
\colhead{($R_{\odot}$)}& 
\colhead{($R_{\odot}$)} & 
\colhead{(K)}& 
\colhead{(K)}& 
\colhead{(days)}& 
\colhead{(d$^{-1}$)}&    
\colhead{}&      
} 

\startdata		

       KIC8262223 &       1.94 & $       0.20$         &      1.67&        1.31&       9128 & 6849& $1.61$&$50-65$&detached  \\
       KIC10661783 &       2.05 & $       0.20$  &      2.56&        1.12  &      7760 & 5980& 1.23&$20-30$&detached    \\
       AS Eri &       1.92 & $      0.21$  &      1.50&        1.15 &      7290 & 4250 & 2.66&$\approx 60$&semi-detached   \\ 
       V228 &       1.51 & $       0.20$  &      1.36&        1.24 &      8070 & 5810 & 1.15&...&semi-detached   \\
\enddata 
\end{deluxetable}

\begin{figure} 
\begin{center} 
{\includegraphics[angle=0,height=12cm]{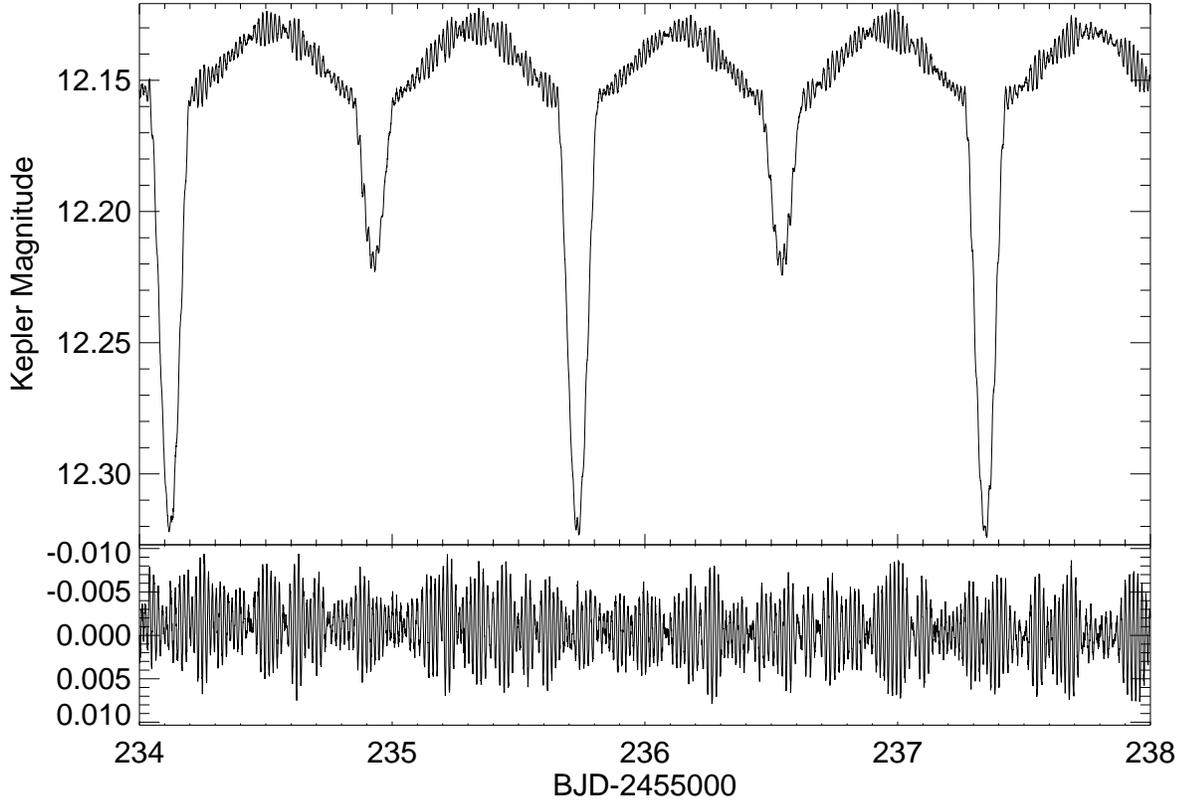}} 
\end{center} 
\caption{The de-trended short cadence light curve of KIC~8262223 during Quarter $4$. The lower panel shows the residuals after subtracting the best binary light curve model.  }
\end{figure}

\begin{figure} 
\begin{center} 
{\includegraphics[angle=0,height=12cm]{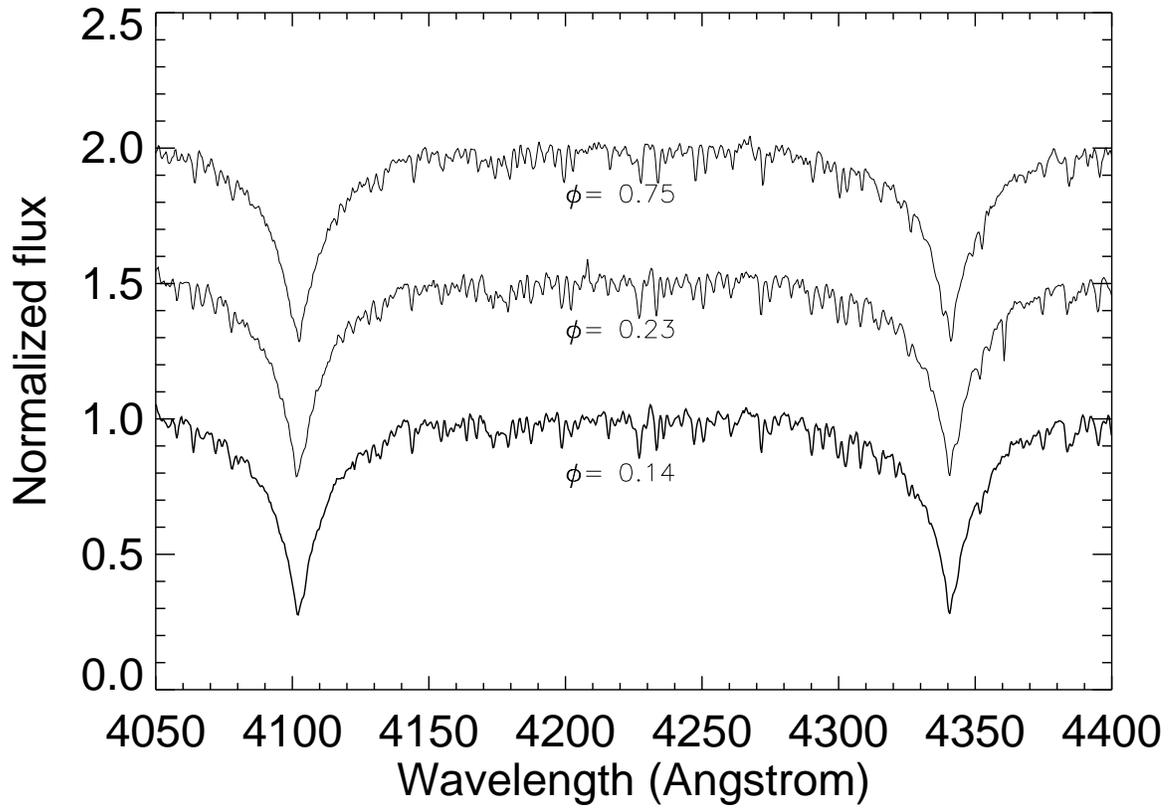}} 
\end{center} 
\caption{The observed composite spectra of KIC~8262223. For clarity, the spectra at orbital phases $\phi=0.23$ and $0.75$ have been shifted upwards by $0.5$ and $1.0$, respectively. The Doppler shifts of the two components can be seen in the core of Balmer lines (H$\delta$ $\lambda 4102$ \AA  \ and H$\gamma$ $\lambda 4341$ \AA). }
\end{figure} 
 
\begin{figure} 
\begin{center} 
{\includegraphics[angle=0,height=12cm]{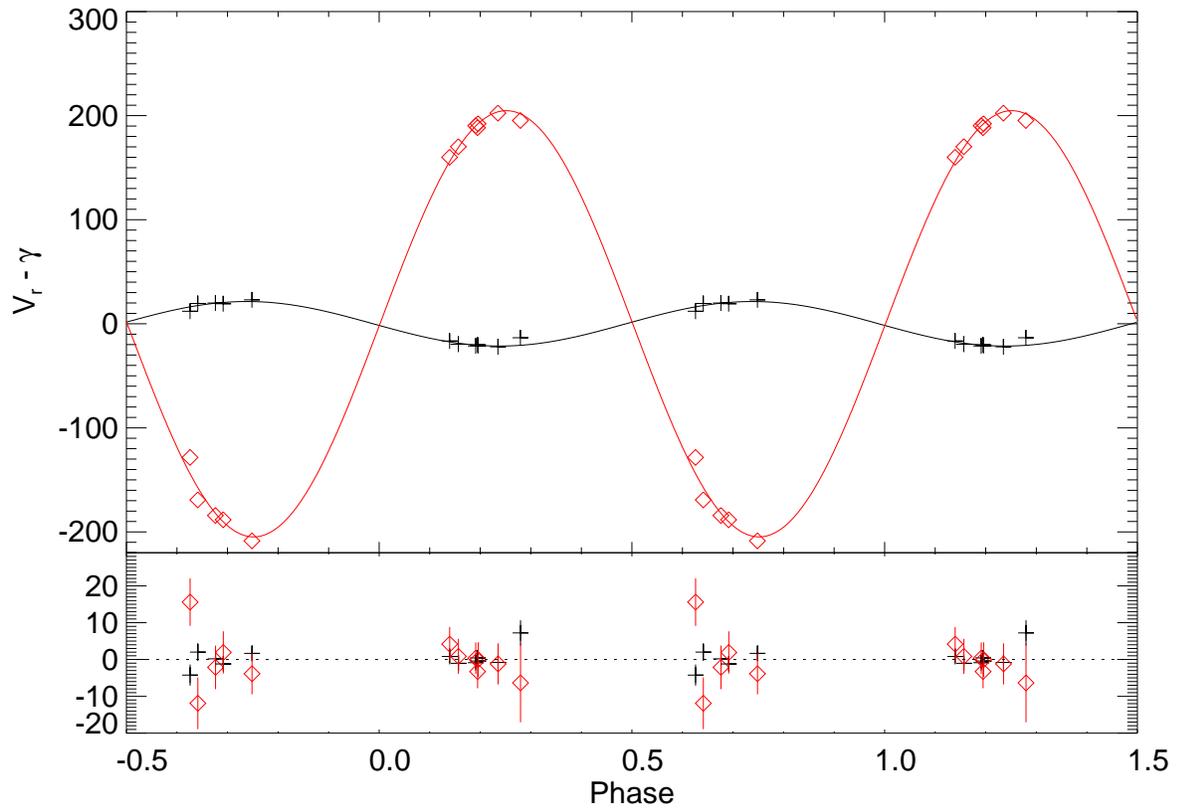}} 
\end{center} 
\caption{Radial velocities (RVs) in km s$^{-1}$ and circular orbital solutions of KIC~8262223. The observed RVs of the primary and the secondary star from cross correlation are shown as black crosses and red diamonds, respectively. The black and red solid lines are the best-fit radial velocity curves for the primary and secondary, respectively. The lower panel shows the corresponding residuals. }
\end{figure}

\begin{figure} 
\begin{center} 
{\includegraphics[angle=0,height=12cm]{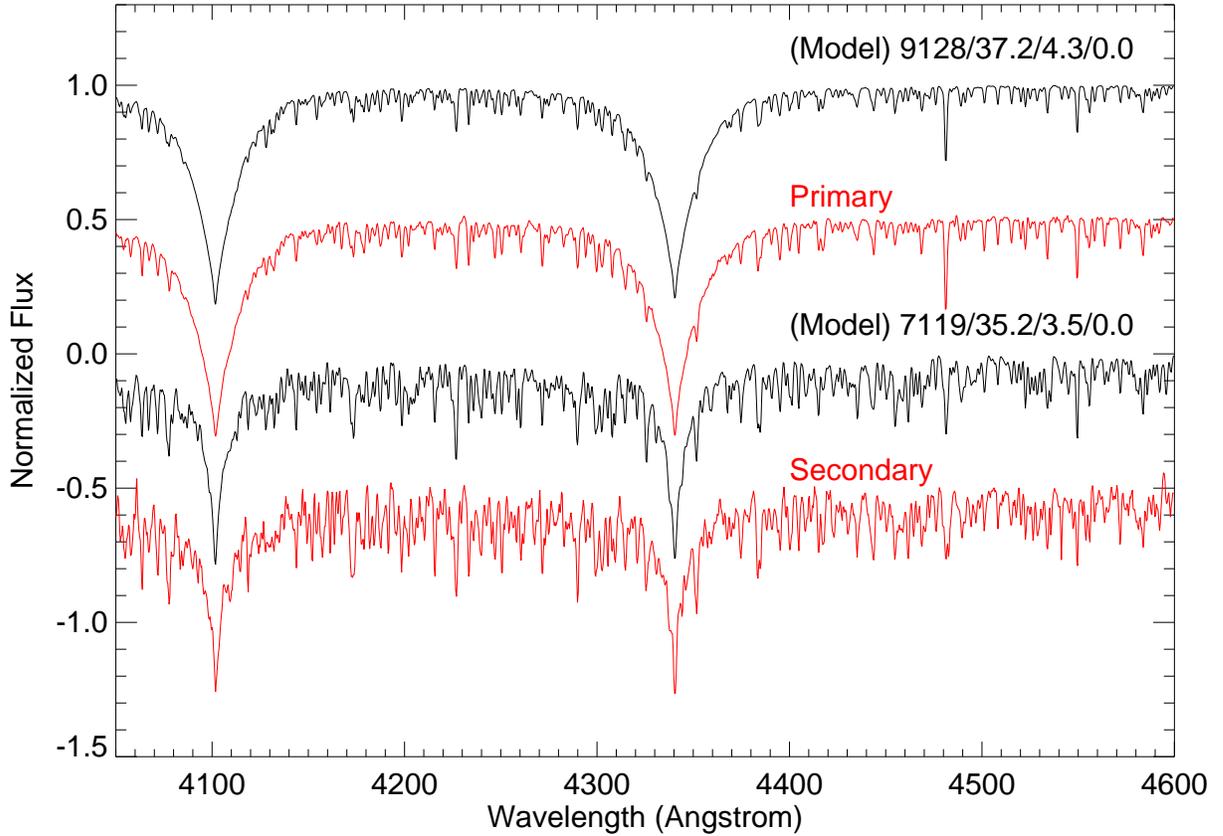}} 
\end{center} 
\caption{The reconstructed individual spectra (red) of the primary (upper) and secondary (lower) of KIC~8262223. The best matching atmospheric models from UVBLUE are shown as black spectra, and the corresponding parameters $T_{\rm eff}$(K), $v \sin i$ (km s$^{-1}$), $\log g$ (cgs) and ${\rm [Fe/H]}$ are labeled. }
\end{figure}




\begin{figure} 
\begin{center} 
{\includegraphics[angle=0,height=12cm]{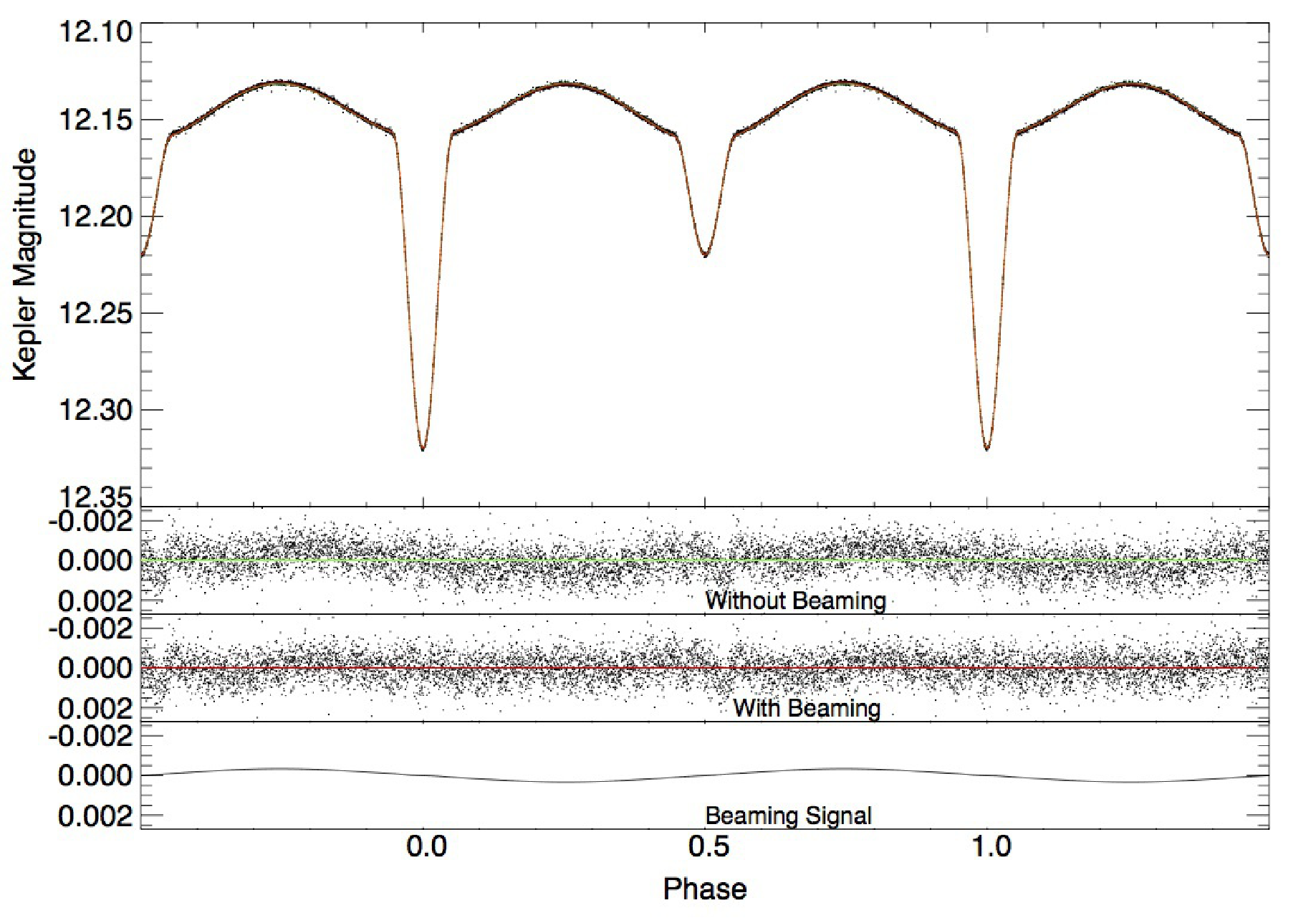}} 
\end{center} 
\caption{\textbf{Top panel}: The phase-folded long cadence light curve of KIC~8262223 (dots) in Quarter 4 and 5 and the best-fit model from ELC (red and green solid line) when the bolometric albedo of the secondary star ($l_2$) is allowed to vary. \textbf{Middle two panels}: The corresponding residuals without and with the Doppler beaming effect taken into account. \textbf{Bottom panel:} ELC model of the beaming lightcurve.}
\end{figure}

\begin{figure} 
\begin{center} 
{\includegraphics[angle=0,height=12cm]{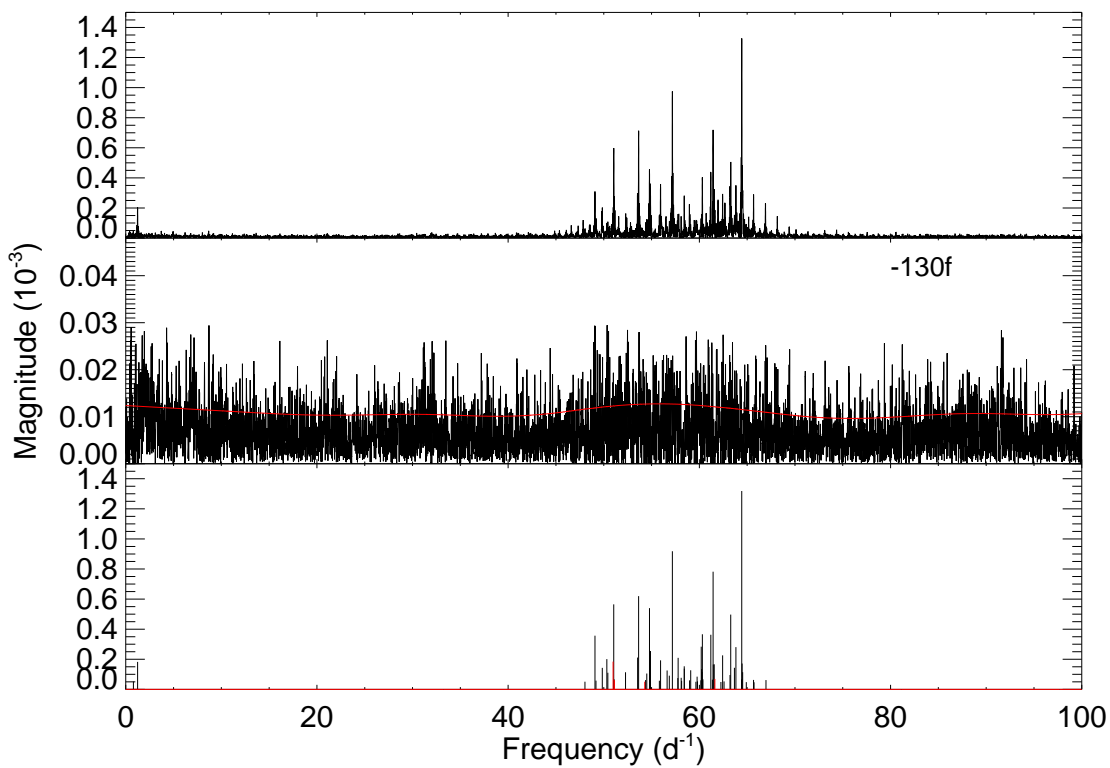}} 
\end{center} 
\caption{\textbf{Top panel:} The amplitude spectrum of the residual light curve of short cadence data (Q$4$) with eclipses masked. \textbf{Middle panel:} The noise spectrum after subtracting $130$ frequencies. The solid red curve represents the adopted noise level which is calcualted by smoothing the envelope of the noise spectrum. \textbf{Bottom panel:} The $64$ extracted significant frequencies with S/N $>$ $4.0$ as listed in Table $5$ (black: independent frequencies; red: combination frequencies).}
\end{figure} 

\begin{figure} 
\begin{center} 
{\includegraphics[angle=0,height=12cm]{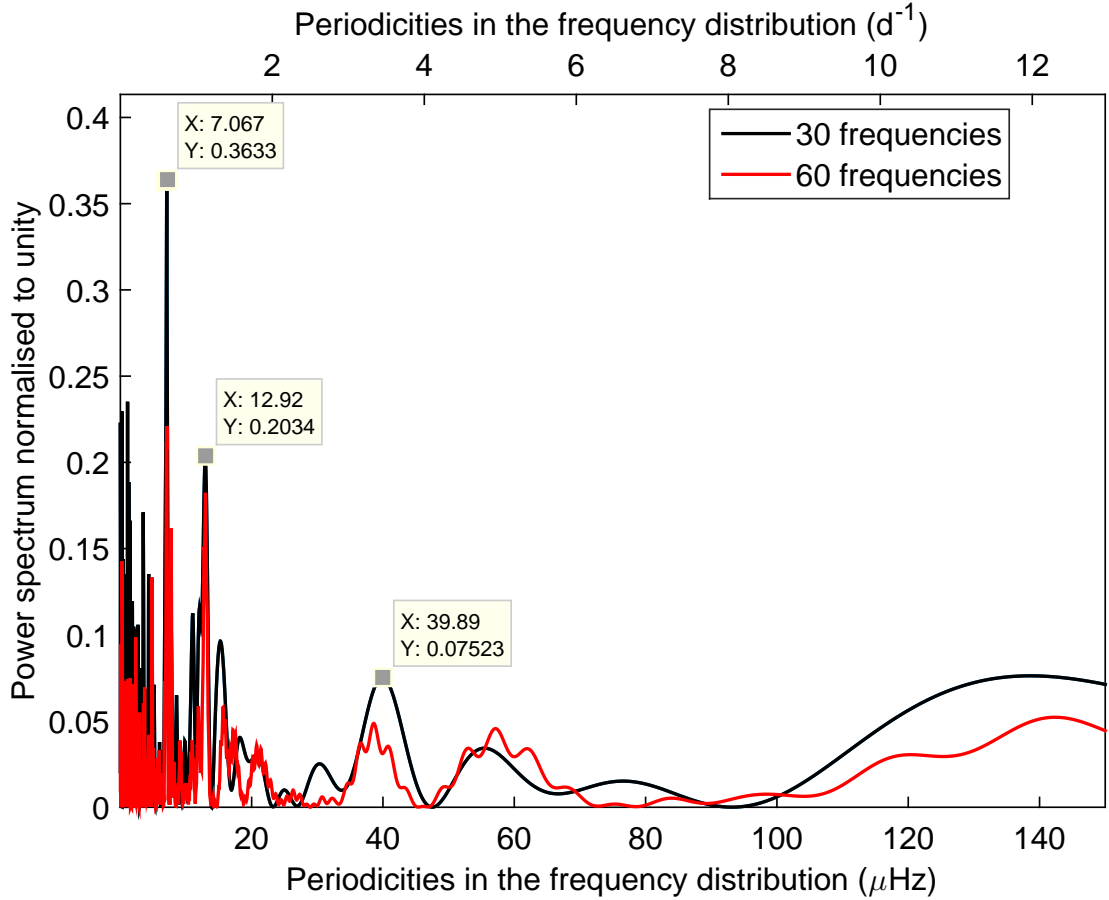}} 
\end{center} 
\caption{The Fourier spectrum of the reported oscillation frequencies in Table 5. The black and red lines are the results of using $30$ and $60$ highest frequencies, respectively. The highest peak at $7.067\mu$HZ (0.61 d$^{-1}$) is likely the result of rotational splitting. The pattern at $39.89\mu$HZ (3.45 d$^{-1}$) is related to half of the large frequency separation $\Delta\nu$. Please see text for more details. }
\end{figure} 

\begin{figure} 
\begin{center} 
{\includegraphics[angle=0,height=12cm]{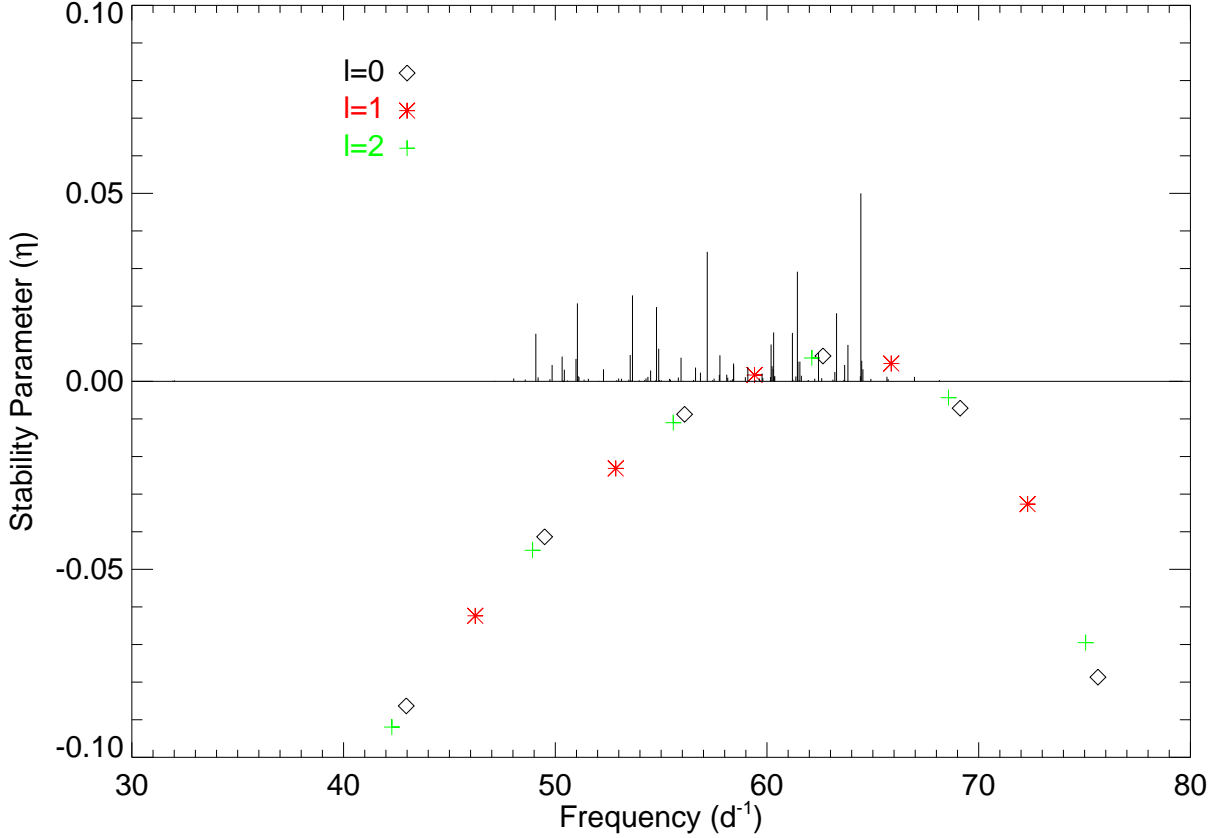}} 
\end{center} 
\caption{The stability parameter $\eta$ of p-modes ($l=0, 1, 2$) of the model for the primary star calculated with MESA and GYRE. The equilibrium model has the following parameters:  $M_1=1.94M_{\odot}$, $R_1=1.67R_{\odot}$, $Z=0.02$, and $Y=0.28$, matching the observed fundamental parameters of KIC 8262223 primary. Unstable modes (positive stability parameter) are in the frequency range $60-67$ d$^{-1}$. The observed frequencies of KIC~8262223 are over-plotted and re-scaled for clarity. }
\end{figure}

\begin{figure} 
\begin{center} 
{\includegraphics[angle=0,height=11cm]{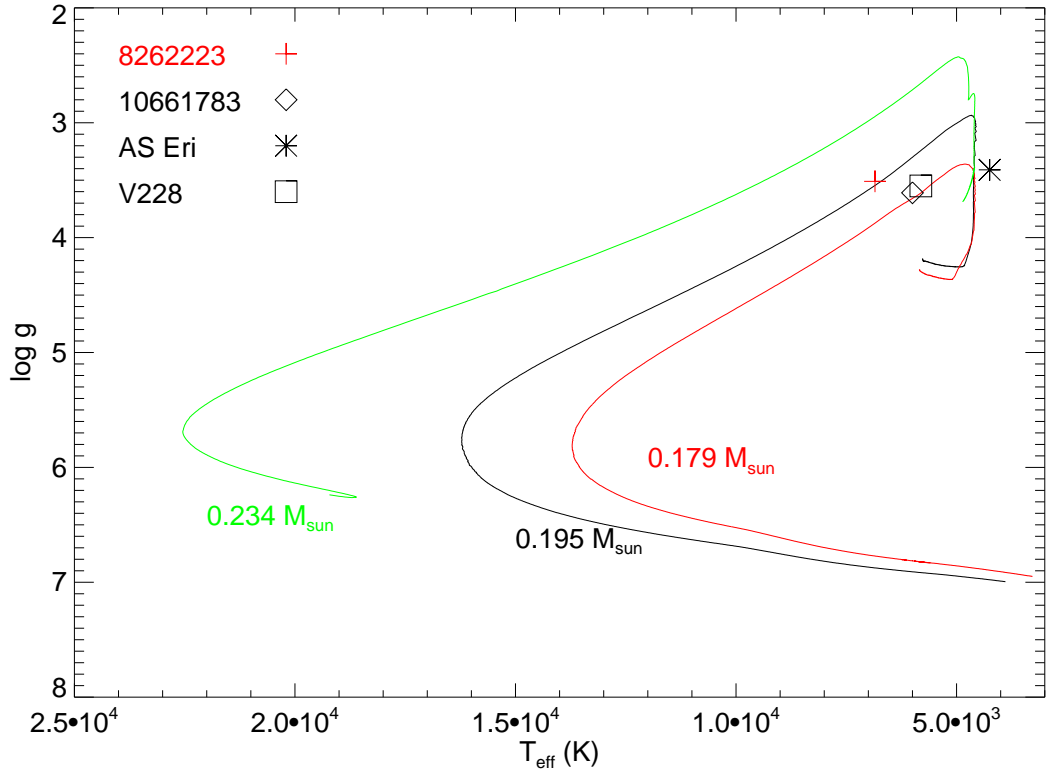}} 
\end{center} 
\caption{The evolutionary tracks of He-WDs calculated by Driebe et al.\ (1999). Three different masses are shown here ($0.179M_{\odot}$, $0.195M_{\odot}$, $0.234M_{\odot}$). The four Algol systems discussed in text are over-plotted.}
\end{figure} 

\begin{figure} 
\begin{center} 
{\includegraphics[angle=0,height=12cm]{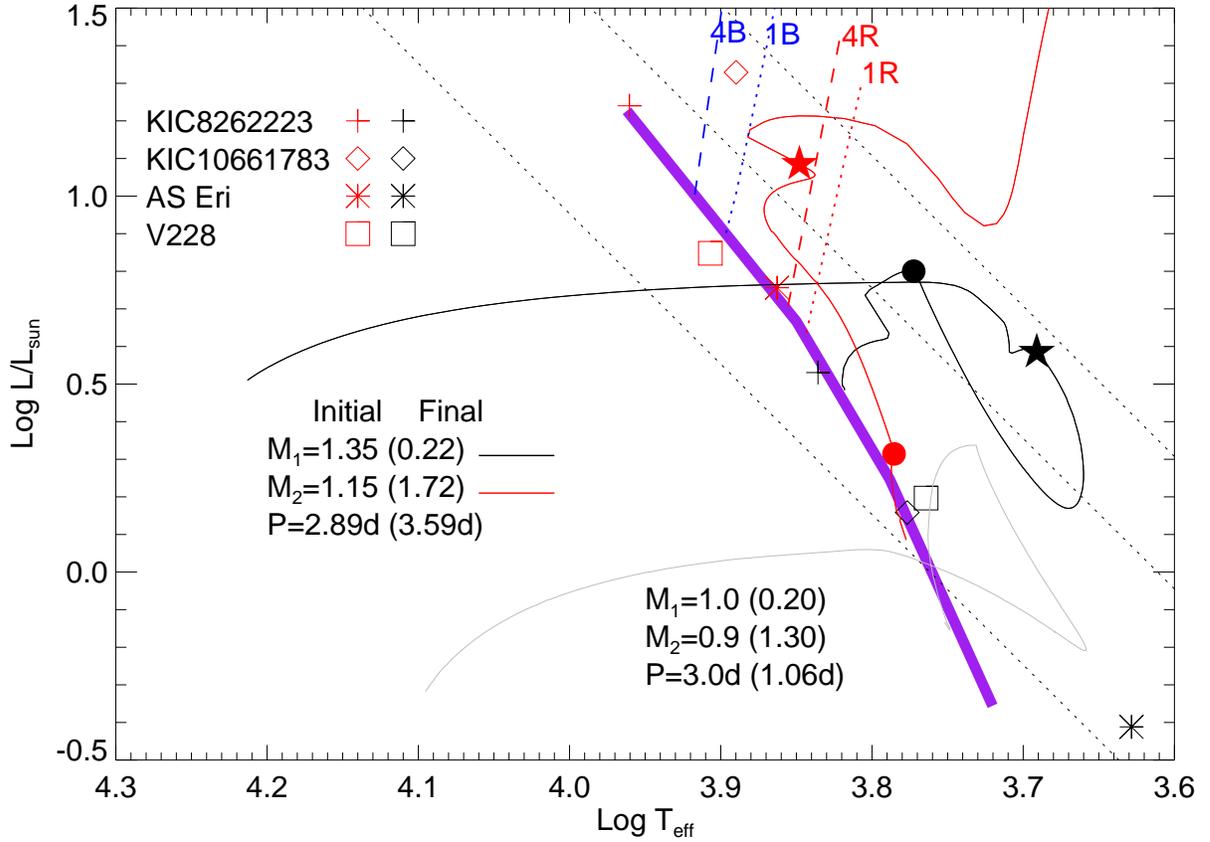}} 
\end{center} 
\caption{Evolutionary tracks for two binary models on the H-R diagram. Model (1) has inital parameters: $M_{10}=1.35M_{\odot},M_{20}=1.15M_{\odot}, 
P=2.89$d and final parameters: $M_{1}=0.22M_{\odot},M_{2}=1.72M_{\odot}, 
P=3.59$d. Model (2) has initial parameters: $M_{10}=1.0M_{\odot},M_{20}=0.9M_{\odot}, 
P=3.0$d and final parameters: $M_{1}=0.20M_{\odot},M_{2}=1.30M_{\odot}, 
P=1.06$d. The evolutionary tracks for the initial primary and secondary of model (1) are shown as red and dark solid lines, respectively. The corresponding track for $M_1$ in model (2) is indicated as the gray line (evolution of $M_2$ not shown). The dashed lines indicate locations of constant radius (from lower left to upper right:  $1R_{\odot}, 2R_{\odot}, 3R_{\odot}$). The locations of four cool Algols in Table 6 are shown as open symbols. Two moments in the evolution are marked for model (1): the onset of mass transfer (filled circle), and the end of mass transfer (filled star). The Zero Age Main Sequence from MESA is indicated by the thick purple line. The blue/red edges of $\delta$ Scuti instability strip calculated by Dupret et al.\ (2005) are indicated by the blue/red lines.}
\end{figure}

\end{document}